\documentclass[utf8]{FrontiersinVancouver} 

\usepackage{url,hyperref,lineno,microtype,subcaption}
\usepackage[onehalfspacing]{setspace}
\usepackage{multirow}
\usepackage{todonotes}

\newcommand{\erw}{\mathbb{E}}
\DeclareMathOperator{\KL}{KL}
\newcommand{\calD}{\mathcal{D}}
\newcommand{\calN}{\mathcal{N}}
\DeclareMathOperator{\ELBO}{ELBO}
\usepackage{bm}

\def\keyFont{\fontsize{8}{11}\helveticabold }
\def\firstAuthorLast{Hort\'ua {et~al.}} 
\def\Authors{H\'ector J. Hort\'ua \,$^{1,2,*}$, Luz \'Angela Garc\'ia\,$^{3}$ and Leonardo Casta\~neda C.\,$^{4}$}
\def\Address{$^{1}$ Grupo Signos, Departamento de Matem\'aticas, Universidad el Bosque, Bogot\'a, Colombia. \\
$^{2}$ Maestr\'ia en Ciencia de Datos, Universidad Escuela Colombiana de Ingeniería Julio Garavito  Bogot\'a, Colombia. \\
$^{3}$ Universidad ECCI. Cra. 19 No. 49-20, Bogot\'a, Colombia, C\'odigo Postal 111311. \\
$^{4}$Observatorio Astron\'omico Nacional, Universidad Nacional de Colombia, Bogot\'a, Colombia. 
}
\def\corrAuthor{-}
\def\corrEmail{hjhortuao@unal.edu.co}

\begin{document}
\onecolumn
\firstpage{1}

\title[Parameter estimation via BNNs]{Constraining cosmological parameters from N-body simulations with Variational Bayesian Neural Networks} 

\author[\firstAuthorLast ]{\Authors} 
\address{} 
\correspondance{} 

\extraAuth{}

\maketitle

\begin{abstract}

\section{}
Methods based on Deep Learning have recently applied  on astrophysical parameter recovery 
thanks to their ability to capture information from complex data. One of these methods are the approximate Bayesian Neural Networks (BNNs) which have demonstrated to yield consistent posterior distribution into the parameter space, helpful for uncertainty quantification. However, as any modern neural networks,  they  tend to produce overly confident uncertainty estimates, and can introduce bias when BNNs are applied to data. In this work, we implement multiplicative normalizing flows (MNFs), a family of approximate posteriors for the parameters of BNNs with the purpose of enhancing the flexibility of the variational posterior distribution, to extract  $\Omega_m$, $h$, and $\sigma_8$ from the QUIJOTE simulations. We have compared this method with respect to the standard BNNs, and the flipout estimator. We found that MNFs combined with BNNs outperform the other models   obtaining  predictive performance with almost one order of magnitude larger that  standard BNNs,   $\sigma_8$  extracted with high accuracy ($r^2=0.99$), and  precise uncertainty estimates. The latter implies  that  MNFs  provide  more realistic predictive distribution closer to the true posterior  mitigating the bias introduced by the variational approximation, and allowing to work with well  calibrated networks.  

\tiny
 \keyFont{ \section{Keywords:} cosmology, N-body simulations, parameter estimation, artificial intelligence, deep neural networks} 
\end{abstract}

\section{Introduction}

Cosmological simulations offer one of the most powerful ways to  understand the initial conditions of the Universe and improve our knowledge on fundamental physics~\citep{2011ASL4204B}. They open also the possibility to fully explore the growth of structure in both the linear and non-linear regime. Currently, the concordance cosmological model, $\Lambda$-CDM, gives an accurate description of most of the observations from early  to late stages of the Universe using a set of few parameters~\citep{dodelson:2003}. Recent observations from Cosmic Microwave Background (CMB)  have provided such accurate estimation for the cosmological parameters, and prompted a tension with respect to local scales measurement, along with  a well-known degeneracy on the total non-relativistic matter density parameters~\citep{2020,Tinker_2011,2204.12180}. Conventionally, the way to capture  information from astronomical observations is to compare summary statistics from data against theory predictions. However, two major difficulties arise: First, it is not well understood what kind of  estimator, or at which  degree of approximation of order statistic should be better  to extract the maximum information from observations. In fact,  the most common choice  is the  power spectrum(PS) which has  shown to be a powerful tool for making inference~\citep{dodelson:2003}. However, It is well known that PS is not able to fully characterize the statistical properties of non-Gaussian density  fields, yielding that  it would not be suitable for  upcoming Large Scale Structure (LSS)  or 21-cm signals  which  are  highly non-Gaussian~\citep{2966.2010.17731,Hamann_2010,Abdalla2022CosmologyIA}. Then, PS  will  miss relevant information  if only this statistic is used for parameter recovery~\citep{2019MNRAS.484..282G}. Second, Cosmologists will require to store and process a large number of data, which can be very expensive. Clearly,  sophisticated computational  tools along with new perspectives on data collection, storage, and analysis  must be developed in order to interpret these observations~\citep{220308056}. \\
In recent years,  artificial intelligence (AI), and Deep  Neural Networks (DNNs)  have emerged as promising tools to tackle the aforementioned difficulties in  the cosmological context  due to its capability for learning relationships between variables in complex data,  outperform traditional estimators, and  handle the demanding computational needs in Astrophysics and Cosmology~\citep{220308056}.
These standard DNNs have been used on a variety of tasks because of their potential for solving inverse problems. However, they are prone to overfitting due to the excessive number of parameters to be adjusted, and the lack of explanations of their predictions for given instances~\citep{Guo:2017:CMN:3305381.3305518}. The latter  is crucial  for cosmological analysis  where  assessing  robustness and reliability of the model predictions are imperative. This problem can be addressed by endowing  DNNs with probabilistic properties that permit quantifying posterior distributions on their outcomes, and  provide them with predictive uncertainties. One of these approaches is  the use of Bayesian Neural Networks (BNNs) comprised  of   probabilistic layers that capture uncertainty over the network parameters (weights), and trained using Bayesian inference~\citep{2106.13594}. Several works have utilized   BNNs in cosmological scenarios where the combination of DNNs (through Convolutional Neural Networks, CNNs) and  probabilistic properties, allow to build models adapted to  non-Gaussian data  without requiring  a priori choice summary statistic~\citep{2019MNRAS.484..282G,1711.02033,Lazanu_2021,2212.06860},  along with quantifying predictive uncertainties~\citep{Hort_a_2020,Hort_a_2020a,2112.11865, PhysRevD.105.023531,List_2020,Wagner_Carena_2021}. 
Indeed, BNNs permit to infer  posterior distributions instead of  point estimates for the weights. These distributions capture the parameter uncertainty, and by subsequently integrating over them, we acquire uncertainties related to the network outputs.  Nevertheless, obtaining the posterior distributions is an intractable task, and approximate techniques such as a Variational Inference(VI)  must be used in order to put them into practice~\citep{NIPS2011_7eb3c8be}. 
Despite the approximate posterior distribution over the weights employed in VI clearly providing  fast computations for inference tasks, they can also introduce a degree of bias  depending on how complex(or simple) the choice of the approximate distribution family is~\citep{2006.01490}. This issue yields overconfident uncertainty predictions and  an unsatisfactory  closeness measurement  with respect to the true posterior. In~\citep{Hort_a_2020a,2112.11865}, the authors included  normalizing flows on the top  of  BNNs to give the joint parameter  distribution  more flexibility. However, that  approach is not implemented into the Bayesian framework, preserving the bias.\\
In this paper, we attempt  to enhance the flexibility of the approximate posterior distribution over the weights of the network by employing multiplicative normalizing flows, resulting in  accurate and precise uncertainty estimates provided by  BNNs. We  apply this approach  to  N-body simulations taken from QUIJOTE dataset~\citep{Villaescusa_Navarro_2020} in order to show  how  BNNs  can take not only advantage of non-Gaussian signals without requiring  a specifying the summary statistic (such as PS)  but also,  increase the posterior complexity, as they yield much larger performance improvements.  
This paper is organized as follows. Section~\ref{secbbn}  
offers a summary of the  BNNs framework and a detailed description of Normalizing flow implementation. Section~\ref{secdata} describes the dataset and analysis tools used in this paper. Numerical implementation and configuration for BNNs are described in Section~\ref{seimplemen}.  
Section~\ref{secanalisis} presents the results we obtained by training BNNs taking into account different approaches and we display the  inference of cosmological parameters. It also outlines the calibration diagrams to determine the accuracy of the uncertainty estimates. Finally, Section~\ref{conlcu} draws the main conclusions of this work and  possible
further directions to the use of BNNs in Cosmology.

\section{Variational Bayesian Neural Networks}\label{secbbn}
Here we go into detail about Bayesian Neural Networks (BNNs), and their implementation  to perform parameter inference. We start with a brief introduction, before focusing on  improving the variational approximation. We remind the reader to refer to~\cite{Abdar_2021,Gal2016Uncertainty,NIPS2011_7eb3c8be} for further details.
\subsection{Approximate BNNs}\label{secbbn1}
The goal of BNNs is to infer the posterior distribution $p(w|\calD)$ over the weights $w$ of the network after observing the data   $\calD=(X, Y)$.  This posterior can be obtained from Bayes law: $p(w|\calD)\sim p(\calD|w)p(w)$, given a likelihood function $p(\calD | w)$, and a prior on the weights $p(w)$. Once the posterior has been computed, the probability distribution on a new test
example $x^*$ is given by
\begin{equation}
  p( y^* |x^*, \calD) = \int_w p(y^*|x^*,w)p(w|\calD)dw,
\end{equation}
where $p(y^*|x^*, w)$ is the predictive distribution for a given value of the weights. For neural networks, however, computing the exact posterior is intractable, so one must resort to approximate BNNs for inference~\citep{Gal2016Uncertainty}.  A popular method to approximate the posterior is variational inference(VI)~\citep{NIPS2011_7eb3c8be}. Let $q(w | \theta)$ be a family of simple distributions parameterized by $\theta$. So, the goal of VI  is to select a distribution $q(w | \theta^*)$ such that $\theta^*$ minimizes $KL\big[q(w|\theta)\big\|p(w|\calD)\big]$, being $KL[\cdot\|\cdot]$  the Kullback-Leibler divergence. This minimization  is equivalent to maximizing the evidence lower bound (ELBO)~\citep{Gal2016Uncertainty}
\begin{equation}\label{elbo}
  \ELBO(\theta) = \erw_{q(w|\theta)}\big[\log p(Y|X,w)\big] - \KL\big[q(w|\theta)\big\|p(w)\big],
\end{equation}
where $\erw_{q(w|\theta)}[\log p(Y|X,w)]$ is the expected log-likelihood with respect to the variational posterior and $\KL[q(w|\theta)||p(w)]$ is the divergence of the variational posterior from the prior. We can observe from Eq.~\ref{elbo} that the KL divergence  acts as a regularizer that encourages the variational posterior moves towards the  modes of the prior.
A common choice for the variational posterior is a product of independent (i.e., mean-field) Gaussian distributions, one distribution for each parameter $w$ in the network~\citep{Abdar_2021}
\begin{equation}\label{meanfield}
    q(w|\theta)=\prod_{ij} \calN(w ;\mu_{ij},\sigma^2_{ij})  
\end{equation}
being $i$ and $j$ the indices of the neurons from the previous layer and the current layer respectively. Applying the reparametrization trick  we arrive at $w_{ij}=\mu_{ij}+\sigma_{ij}*\epsilon_{ij}$, where $\epsilon_{ij}$ is drawn from a standard normal distribution. Furthermore, if the prior is also a product of independent Gaussians, the KL divergence between the prior and the variational posterior be computed analytically, which makes this approach  computationally efficient. 
\subsubsection{Flipout}\label{secbbn2}
In case where  sampling from $q(w|\theta)$ is not  fully independently for different examples in a mini-batch, we well obtain  gradient estimates with high variance. Flipout method provides an alternative to decorrelate the gradients within a mini batch by implicitly sampling pseudo-independent weights for each example~\cite{1803.04386}. The method  requires  two assumptions about the properties of $q(w|\theta)$: symmetric with respect to zero, and the weights of the network are independent. Under these assumptions, the distribution is invariant to element wise multiplication by
a random sign matrix $\hat{r}$, i.e., $\hat{w}=w \circ \hat{r}$, implies that $w\sim q(w)\approx\hat{w}\sim \hat{q}(\hat{w})$. Therefore, the marginal distribution over gradients computed for individual examples will be identical to the distribution computed using shared weights samples. Hence, Flipout achieves much lower variance updates when averaging over a mini batch. We validate this approach experimentally by comparing against Multiplicative normalizing flows.
\subsection{Uncertainty in BNNs}\label{secbbn3}
BNNs offer a groundwork  to incorporate from the posterior distribution  both, the uncertainty inherent to the data (aleatoric uncertainty), and the uncertainty in the model parameters due to a limited amount of training data (epistemic uncertainty)~\citep{KIUREGHIAN2009105}. Following~\citep{Hort_a_2020}, assuming that the top of the BNNs  consist of a mean vector $\mu\in\mathbb{R}^{N}$ and a covariance matrix $\Sigma\in\mathbb{R}^{N(N+1)/2}$\footnote{Where the targets $y \in \mathbb{R}^N$.}, and for a given fixed input $x^*$, $T$ forward passes of the network are computed, obtaining for each of their mean  $\mu_t$ and  covariance matrix $\Sigma_t$. Then,  an estimator for approximate the predictive covariance  can be written as  
\begin{equation}\label{eq:7}
\widehat{\mathrm{Cov}}(y^*,y^*|x^*)\approx \underbrace{\frac{1}{T}\sum_{t=1}^{T}\Sigma_t}_\text{Aleatoric}+ \underbrace{\frac{1}{T}\sum_{t=1}^{T}( {{\mu}}_{t}-\overline{\mu})( {\mu}_{t}-\overline{\mu})^{\mathrm{T}}}_\text{Epistemic}, 
\end{equation}
with $\overline{\mu}= \frac{1}{T}\sum_{t=1}^{T} {\mu}_t$.
 Notice that in case $\Sigma$ is diagonal, and $\sigma^2 = \text{diag}(\Sigma)$, the last equation reduces to the results obtained  in~\cite{kendall2017uncertainties,kwon} 
\begin{equation}\label{eq:8}
\widehat{\mathrm{Var}}(  y^*|x^*)\approx \underbrace{\frac{1}{T}\sum_{t=1}^{T} \sigma^2_{t}}_\text{Aleatoric}+ \underbrace{\frac{1}{T}\sum_{t=1}^{T}{(\mu}_{t}-\bar{\mu})^2}_\text{Epistemic}. 
\end{equation}
In this scenario, BNNs can be used to learn the correlations between the the targets and produce estimates of their uncertainties. Unfortunately, the uncertainty computed from Eqs.~\ref{eq:7},~\ref{eq:8}, tends to be miscalibrated, i.e., the predicted uncertainty (taking into account both epistemic and aleatoric uncertainty) is underestimated and does not allow robust detection of uncertain predictions at inference. Therefore, calibration diagrams along with methods to jointly calibrate aleatoric and epistemic uncertainties, must be employed before inferring predictions from BNNs~\cite{laves2020wellcalibrated}. We come back to this point in Section 5.

\subsection{Multiplicative normalizing flows}\label{secbbn4}
As mentioned previously, the most common family for the variational posterior used in BNNs is the mean-field Gaussian distributions defined in Eq.~\ref{meanfield}. This simple
distribution  is unable to capture the complexity of the true posterior. Therefore, we expect that increasing the complexity of the variational posterior, BNNs  achieve significant
performance gains  since we are now able to sample from a complicate distribution that more closely resembles the true posterior. Certainly, transforming the variational posterior must be followed with   fast computations   and still being numerically tractable.  We now describe in detail the  Multiplicative Normalizing Flows (MNFs) method that  provides flexible posterior distributions in an efficient way by employing auxiliary random variables and normalizing flows proposed by~\citep{10.5555/3305890.3305910}. MNFs  propose that the variational posterior can be expressed as an infinite mixture of distributions
\begin{equation}\label{mixture}
    q(w|\theta)= \int q(w|z,\theta)q(z|\theta)dz 
\end{equation}
where $\theta$ is the learnable posterior parameter, and $z\sim q(z|\theta)\equiv q(z)$\footnote{The parameter $\theta$ will be omitted in this section for clarity of notation.} is a vector with the same dimension on the input layer,  which  plays the role of an auxiliary latent variable. Moreover, allowing local reparametrizations, the variational posterior for fully connected layers become a modification of  Eq.~\ref{meanfield} written as
\begin{equation}\label{mnfpos}
   w \sim q(w|z)=\prod_{ij}\calN(w;z_i\mu_{ij},\sigma^2_{ij}).
\end{equation}
Notice that by enhancing the complexity of $q(z)$, we can increase the flexibility
of the variational posterior. This  can be done using Normalizing Flows since the dimensionality of $z$ is much lower compared to the weights. Starting from samples $z_0\sim q(z_0)$ from   fully factorized Gaussian Eq.~\ref{meanfield}, a rich distribution $q(z_K)$ can be obtained by applying a successively invertible K-transformations $f_K$ on $z_0$
\begin{equation}\label{nfz0}
    z_K=\text{NF}(z_0)=f_K \circ\cdots\circ f_1(z_0); \quad \log q(z_K)=\log q(z_0) - \sum_{k=1}^K\log \left|\det\frac{\partial f_k}{\partial z_{k-1}} \right|.
\end{equation}
Unfortunately, the KL divergence in  Eq.~\ref{elbo}  becomes generally intractable as the posterior $q(w)$ is an infinite mixture as shown in Eq.~\ref{mixture}. This is addressed also in~\citep{3045661} by  evoking Bayes law $q(z_K)q(w|z_K)=q(w)q(z_K|w)$ and  introducing an auxiliary  distribution $r(z_K | w,\phi)$ parameterized by $\phi$, with the purpose of approximating the  posterior distribution of the original variational parameters $q(z_K|w)$  to further lower bound the KL divergence term. Therefore,  KL divergence term  can be bounded as follows
\begin{equation}\label{velbo}
\begin{split}
      - \KL\big[q(w)\big\|p(w)\big] &=-\erw_{q(w)}\left[\log \left(\frac{q(w)}{p(w)} \right) \right]\\
       & \geq -\erw_{q(w)}\left[\log \left(\frac{q(w)}{p(w)} \right) +\KL\big[q(z_K|w)\big\|r(z_K|w,\phi)\big]\right]\\
        & = -\erw_{q(w)}\left[\log \left(\frac{q(w)}{p(w)} \right) +\erw_{q(z_K|w)}\left[\log \left(\frac{q(z_K|w)}{r(z_K|w,\phi)} \right) \right]\right] \\
          & = -\erw_{q(w)}\left[ \erw_{q(z_K|w)}\left[\log \left(\frac{q(w)}{p(w)} \right) \right] +\erw_{q(z_K|w)}\left[\log \left(\frac{q(z_K|w)}{r(z_K|w,\phi)} \right) \right]\right] \\
              & = -\erw_{q(w,z_K)}\left[ \log \left(\frac{q(w)}{p(w)} \right)  +\log \left(\frac{q(z_K|w)}{r(z_K|w,\phi)} \right) \right] \\
                & = \erw_{q(w,z_K)}\left[ -\log \left(q(w)q(z_K|w)\right)  +\log r(z_K|w,\phi)   +\log p(w)  \right]  \Rightarrow\\
 - \KL\big[q(w)\big\|p(w)\big] &\geq \erw_{q(w,z_K)}\left[ - \KL\big[q(w|z_K)\big\|p(w)\big]+\log q(z_K)  +\log r(z_K|w,\phi)     \right], \\
\end{split}
\end{equation}
where we have taken into account that $\text{KL}[P\|Q]\geq 0$, and the equality is satisfied iff $P=Q$.  In the last line, the first term   can be analytically computed since it will  be the KL divergence between two Gaussian distributions, while  the second term is given by the Normalizing flow generated by $f_K$ as we observe in Eq.~\ref{nfz0}. Finally, the auxiliary posterior term is parameterized by inverse normalizing flows as follows~\cite{1806.02315}
\begin{equation}\label{nfz1}
    z_0=\text{NF}^{-1}(z_K)=g^{-1}_1 \circ\cdots\circ g^{-1}_K(z_K); \quad \log r(z_K|w,\phi)=\log r(z_0|w,\phi) + \sum_{k=1}^K\log \left|\det\frac{\partial g^{-1}_k}{\partial z_{k}} \right|,
\end{equation}
where one can  parameterize $g^{-1}_K$ as another normalizing flow. In the paper~\citep{10.5555/3305890.3305910}, the authors  also propose a flexible parametrization  of the auxiliary posterior as 
\begin{equation}
    z_0\sim  r(z_K|w,\phi)=\prod_i \calN(z_0;\tilde{\mu}_i(w,\phi),\tilde{\sigma}^2_i(w,\phi)).
\end{equation}
We will use the parameterization of  the mean  $\tilde{\mu}$, and the variance $\tilde{\sigma}^2$  as in the original paper as well as the masked RealNVP~\citep{dinh2017density} as choice of Normalizing flows. 

\section{N-body simulations dataset}\label{secdata}
In this work, we leverage 2000 hypercubes simulation taken from The \href{https://quijote-simulations.readthedocs.io/en/latest/}{ Quijote project}~\citep{Villaescusa_Navarro_2020}. They have been run using the TreePM
code Gadget-III~\citep{Springel_2005}, and their  initial conditions were generated at $z=127$ using 2LPT~\cite{Scoccimarro_1998}.
The set chosen for this work is made of standard simulations with different random seeds  with the intention of emulating  the cosmic variance. Each instance corresponds to  a three-dimensional distribution of the density field with size  $64^3$.   The cosmological parameters vary  according to $\Omega_m \in [0.1,0.5]$, $\Omega_b \in [0.03,0.07]$, $h \in [0.5,0.9]$, $n_s \in [0.8,1.2]$, $\sigma_8 \in [0.6,1.0]$, while  neutrino mass ($M_{\nu}=0$eV) and the equation of state parameter ($w=-1$) are  kept fixed.  The dataset was split into training($70\%$), validation ($10\%$), and test ($20\%$), while   hypercubes were logarithmic transformed and the cosmological  parameters  normalized between 0 and 1.
In this paper we will build BNNs with the ability to predict  three out of five aforementioned parameters, $\Omega_m$, $\sigma_8$ and $h$.

\section{BNNs Implementation}\label{seimplemen}
 We will consider three different BNNs architectures
based on the discussion presented in Section~\ref{secbbn}: standard BNNs (prior and variational posterior defined as a mean-field Normal distributions) \textbf{[sBNNs]}; BNNs with Flipout estimator \textbf{[FlipoutBNNs]}; and Multiplicative normalizing flows \textbf{[VBNNs]}. The experiments were implemented using the \href{https://www.tensorflow.org/}{TensorFlow v:2.9} and \href{https://www.tensorflow.org/probability}{TensorFlow-probability v:0.19}~\citep{tensorflow2015-whitepaper}. All  BNNs  designed in this paper are comprised of three parts. First, all experiments start with a $64^3$-voxel input layer corresponding to the normalised 3D density field   followed by  the fully-convolutional ResNet-18 backbone as it is presented schematically in table~\ref{tab:BNN_}.  All the Resblock are fully pre-activated and their representation can be seen in  figure.~\ref{fig: subfigures}. The repository \href{https://github.com/ZFTurbo/classification_models_3D}{Classification models 3D} was used to build the backbone of BNNs~\citep{solovyev20223d}. Subsequently, the second part  of BNNs represents the stochasticity of the network. This is comprised of just one layer and it depends on the type of BNN used. For \textit{sBNNs}, we employ  the dense variational layer which   uses variational inference to fit an approximate  posterior to the distribution over both the kernel matrix and the bias terms. Here, we use as posterior and  prior(no-trainable)   Normal distributions.  Experiments with \textit{FlipoutBNNs} for instance, are made via Flipout dense layer where the mean field normal distribution are also utilized  to parameterize the distributions. These two layers are already implemented in the package  \href{https://www.tensorflow.org/probability}{TF-probability}~\citep{tensorflow2015-whitepaper}. On the other hand, for  VBNNs we have adapted the class DenseMNF implemented in the repositories  
 \href{https://github.com/janosh/tf-mnf/tree/0ed492bd8faf0bdc37a56f87adf2d8ca425eec5b}{TF-MNF}, \href{https://github.com/AMLab-Amsterdam/MNF_VBNN}{MNF-VBNN}~\citep{10.5555/3305890.3305910} to our model. Here, we use $50$ layers for the masked RealNVP NF, and the maximum  variance for layer weights is around the unity. Finally, the last part of  all BNNs  account for the  output of the network, which is dependent on the aleatoric uncertainty parameterization. We use a 3D multivariate Gaussian distribution with nine parameters to be learnt (three means $\mathbf{\mu}$ for the cosmological parameters, and six elements for the covariance matrix $\Sigma$). 

\begin{table}
\tiny
\centering
\begin{tabular}{ |p{4cm}||p{4cm}||p{4cm}| }
\hline
 \multicolumn{3}{|c|}{ \textbf{ResNet-18 backbone} } \\ [0.1cm]
 \hline
\textbf{Layer Name} & \textbf{Input Shape} & \textbf{Output Shape} \\ [0.1cm] \hline
Batch Norm & ($N_\text{batch}$, 64,64,64,3) & ($N_\text{batch}$, 64,64,64,3) \\ [0.1cm]
3D Convolutional & ($N_\text{batch}$, 70,70,70,3) & ($N_\text{batch}$, 32,32,32,64) \\  [0.1cm]
Batch Norm+ReLU & ($N_\text{batch}$, 32,32,32,64) & ($N_\text{batch}$, 32,32,32,64) \\  [0.1cm]
Max Pooling 3D & ($N_\text{batch}$, 34,34,34,64) & ($N_\text{batch}$, 16,16,16,64) \\  [0.1cm]
Batch Norm+ReLU & ($N_\text{batch}$,  16,16,16,64) & ($N_\text{batch}$,  16,16,16,64) \\ [0.1cm]
Resblock 1& $\begin{bmatrix}
(N_\text{batch},  16,16,16,64) \\
(N_\text{batch},  16,16,16,64) \end{bmatrix}$  & ($N_\text{batch}$,  16,16,16,64)\\ [0.5cm]
Batch Norm+ReLU & ($N_\text{batch}$,  16,16,16,64) & ($N_\text{batch}$,  16,16,16,64) \\  [0.3cm]
Resblock 2& $\begin{bmatrix}
(N_\text{batch},  16,16,16,64) \\
(N_\text{batch},  8,8,8,128) \end{bmatrix}$  & ($N_\text{batch}$,  8,8,8,128)\\ [0.5cm]
Batch Norm+ReLU & ($N_\text{batch}$, 8,8,8,128 ) & ($N_\text{batch}$,  8,8,8,128) \\ [0.3cm] 
Resblock 3& $\begin{bmatrix}
(N_\text{batch},   8,8,8,128) \\
(N_\text{batch},  4,4,4,256) \end{bmatrix}$  & ($N_\text{batch}$,  4,4,4,256)\\[0.5cm]
Batch Norm+ReLU & ($N_\text{batch}$, 4,4,4,256 ) & ($N_\text{batch}$,  4,4,4,256) \\  [0.3cm]
Resblock 4& $\begin{bmatrix}
(N_\text{batch},   4,4,4,256) \\
(N_\text{batch},  2,2,2,512) \end{bmatrix}$  & ($N_\text{batch}$,  2,2,2,512)\\ [0.5cm]
Batch Norm+ReLU & ($N_\text{batch}$, 2,2,2,512 ) & ($N_\text{batch}$,   2,2,2,512) \\  [0.1cm]
Global Avg Pooling & ($N_\text{batch}$, 2,2,2,512) & ($N_\text{batch}$, 512) \\ 
\hline
\end{tabular}
\caption{Configuration of the backbone BNNs used for all experiments presented in this paper.}
\label{tab:BNN_}
\end{table}

\setcounter{figure}{1}
\setcounter{subfigure}{0}
\begin{subfigure}
\setcounter{figure}{1}
\setcounter{subfigure}{0}
    \centering
    \begin{minipage}[b]{0.5\textwidth}
        \includegraphics[width=\linewidth]{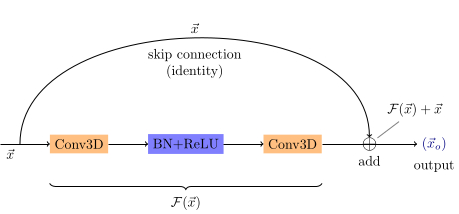}
        \caption{Illustration of the first skip connection in a residual block.}
        \label{fig:Subfigure 1}
    \end{minipage}  
   
\setcounter{figure}{1}
\setcounter{subfigure}{1}
    \begin{minipage}[b]{0.5\textwidth}
        \includegraphics[width=\linewidth]{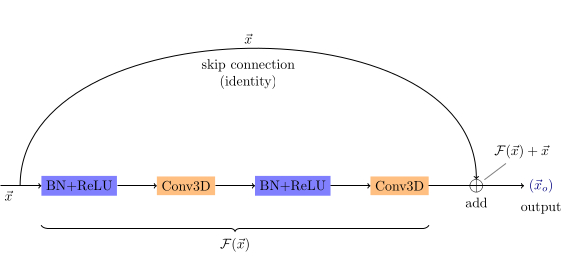}
        \caption{Illustration of the second skip connection in the residual block.}
        \label{fig:Subfigure 2}
    \end{minipage}

\setcounter{figure}{1}
\setcounter{subfigure}{-1}
    \caption{Each Resblock includes both skip connection configurations. \textbf{(A)} The Resblock starts with this configuration applied to the input tensor. \textbf{(B)} The output of  the previous configuration is fed into this connection.}
    \label{fig: subfigures}
\end{subfigure}

 The loss function to be optimized during training is given by the ELBO~\ref{elbo} where the second term is associated to the negative log-likelihood (NLL) 
 \begin{equation}\label{nll}
     -\text{NLL}\sim \frac{1}{2}\log|s\cdot\Sigma|+\frac{1}{2}(\mathbf{y}-\mathbf{\mu})^\top
     \left(s\cdot\Sigma\right)^{-1} (\mathbf{y}-\mathbf{\mu}),
 \end{equation}
averaged over the mini-batch. The scalar variable $s$ is equal to one during the training process, and it becomes a trainable variable during post-training  to recalibrate the probability density function~\citep{Hort_a_2020,laves2020wellcalibrated}. The algorithm used to minimize the objective function is the Adam optimizer  with first and second moment
exponential decay rates of 0.9 and 0.999, respectively~\cite{https://doi.org/10.48550/arxiv.1412.6980}. The
learning rate starts from $10^{-3}$ and it will be reduced by a factor of $0.8$ in case that any improvement has not been observed after 10 epochs. Furthermore, we have applied  warm-up period for which the model turns on progressively the KL term in Eq.~\ref{elbo}. This is achieved by introducing a $\beta$ variable in the ELBO, i.e., $\beta\cdot\KL\big[q(w|\theta)\big\|p(w)\big]$, so,  this parameter  starts being equal to 0 and grows linearly to 1 during 10 epochs~\cite{3157382.3157516}. BNNs were trained with 32 batches and early stopping callback for avoiding over-fitting. The infrastructure used was  the Google Cloud Platform \href{https://cloud.google.com/}{(GCP)} using a nvidia-tesla-t4 of  16 GB GDDR6 in a N1 machine series shared-core.     
\subsection{Metrics}
We compare all BNN results  in terms of performance, i.e., the precision of their predictions for the cosmological parameters quantified through Mean Square Error (MSE),  ELBO, and  plotting the true vs predicted values with  its coefficient of determination.  Also, it is important to quantify the quality of the uncertainty estimates. One of the ways to diagnostic the quality of the uncertainty estimates is through reliability diagrams. Following~\cite{laves2020wellcalibrated,Guo:2017:CMN:3305381.3305518},   we can define perfect calibration of regression uncertainty as


\begin{equation}
    \mathbb{E}_{\hat{\sigma}^{2}} \left[ \big| \big( \mathbb{E}[(\bm{y}-\bm{\mu})^{2}] \, \big| \, \hat{\sigma}^{2} = \alpha^{2} \big) - \alpha^{2} \big| \right] \quad \forall \left\{ \alpha^{2} \in \mathbb{R} \, \big| \, \alpha^{2} \geq 0 \right\}.
    \label{eq:uce}
\end{equation}
Hence, the predicted uncertainty  $ \hat{\sigma}^{2} $  is partitioned into $ K $ bins with equal width, and the  variance per bin is defined as
\begin{equation}
    \mathrm{var}(B_{k}) := \frac{1}{\big| B_{k} \big|} \sum_{i \in B_{m}} {\frac{1}{N} \sum_{n=1}^{N} \left( \bm{\mu}_{i,n} - \bm{y}_{i} \right)^{2} } ,
\end{equation}
with $ N $ stochastic forward passes. On the other hand,  the uncertainty per bin is defined as
\begin{equation}
    \mathrm{uncert}(B_{k}) := \frac{1}{\vert B_{k} \vert} \sum_{i \in B_{k}} \hat{\sigma}_{i}^{2}.
\end{equation}
With these two quantities, we can generate reliability diagrams to assess
the quality of the estimated uncertainty via  plotting $ \mathrm{var}(B_{k}) $ vs.\ $ \mathrm{uncert}(B_{k}) $. In addition, we can compute the expected uncertainty calibration error (UCE) in order to quantify  the miscalibration 
\begin{equation}\label{uce}
    \mathrm{UCE} := {\sum_{k=1}^{K}} \frac{\vert B_{k} \vert}{m} \big| {\mathrm{var}}(B_{k}) - \mathrm{uncert}(B_{k}) \big| ,
\end{equation}
with number of inputs $m$ and set of indices $ B_{k} $ of inputs, for which the uncertainty falls into the bin $ k $. A more general approach proposed in~\cite{Hort_a_2020} consists in computing the expected coverage probabilities defined as the x\% of samples for which the
true value of the parameters falls in the x\%-confidence region defined by the joint posterior. Clearly, this option is more precise since it captures  higher-order statistics  through the full posterior distribution. However, for simplicity, we will follow the UCE approach.

\section{Analysis and Results of  parameter inference with BNNs}\label{secanalisis}
In this section we discuss the results obtained by comparing  three different versions of BNNs, the one with MNFs, the standard BNN, and the third one using  Flipout as estimator. The results reported in this section were computed on  the Test dataset. Table~\ref{table2}. shows the metrics obtained for each BNN approach. As mentioned, MSE, ELBO and $r^2$ provide well estimates for determining the precision of the model, while UCE measures  the miscalibration.  Here, we can observe that VBNNs outperform all experiments, not only taking into account the average error, but also the precision for each cosmological parameter along with a good calibration in its uncertainty predictions. 
\begin{table}[]
\resizebox{\linewidth}{!}{%
\begin{tabular}{|l|lllll|lllll|lllll|}
\hline
Metrics & \multicolumn{5}{l|}{FlipoutBNNs}                                                                                                             & \multicolumn{5}{l|}{VBNNs}                                                                                                                  & \multicolumn{5}{l|}{sBNNs}                                                                                                             \\ \hline
        & \multicolumn{1}{l|}{$\Omega_m$}    & \multicolumn{1}{l|}{$\sigma_8$}    & \multicolumn{1}{l|}{$h$}    & \multicolumn{1}{l|}{$\Omega_m h^2$} & $\sigma_8\Omega_m^{0.25}$   & \multicolumn{1}{l|}{$\Omega_m$}    & \multicolumn{1}{l|}{$\sigma_8$}    & \multicolumn{1}{l|}{$h$}    & \multicolumn{1}{l|}{$\Omega_m h^2$} & $\sigma_8\Omega_m^{0.25}$  &  \multicolumn{1}{l|}{$\Omega_m$}    & \multicolumn{1}{l|}{$\sigma_8$}    & \multicolumn{1}{l|}{$h$}    & \multicolumn{1}{l|}{$\Omega_m h^2$} & $\sigma_8\Omega_m^{0.25}$  \\ \hline
MSE     & \multicolumn{3}{l|}{0.063}                                                        & \multicolumn{2}{l|}{}                             & \multicolumn{3}{l|}{0.057}                                                            & \multicolumn{2}{l|}{}                             & \multicolumn{3}{l|}{0.190}                                                        & \multicolumn{2}{l|}{}                             \\ \hline
ELBO    & \multicolumn{3}{l|}{20.85}                                                        & \multicolumn{2}{l|}{}                             & \multicolumn{3}{l|}{19.71}                                                            & \multicolumn{2}{l|}{}                             & \multicolumn{3}{l|}{31.57}                                                        & \multicolumn{2}{l|}{}                             \\ \hline
$r^2$      & \multicolumn{1}{l|}{0.82} & \multicolumn{1}{l|}{0.98} & \multicolumn{1}{l|}{0.2}  & \multicolumn{1}{l|}{0.03}                  & 0.93 & \multicolumn{1}{l|}{0.85}   & \multicolumn{1}{l|}{0.99}  & \multicolumn{1}{l|}{0.4}   & \multicolumn{1}{l|}{0.56}                  & 0.95 & \multicolumn{1}{l|}{0.75} & \multicolumn{1}{l|}{0.85} & \multicolumn{1}{l|}{0.01} & \multicolumn{1}{l|}{0.23}                  & 0.80 \\ \hline
 UCE   & \multicolumn{1}{l|}{0.109} & \multicolumn{1}{l|}{8.10}  & \multicolumn{1}{l|}{0.26} & \multicolumn{2}{l|}{}                             & \multicolumn{1}{l|}{0.0008} & \multicolumn{1}{l|}{0.0008} & \multicolumn{1}{l|}{0.010} & \multicolumn{2}{l|}{}                             & \multicolumn{3}{l|}{\textgreater{}1.0}                                            & \multicolumn{2}{l|}{}                             \\ \hline
\end{tabular} 

}\caption{Metrics test set results for all BNNs architectures. High UCE values indicate miscalibration.  MSE and ELBO are computed only over the cosmological parameters. }\label{table2}
\end{table}
\begin{figure}[h!]
\begin{center}
\includegraphics[width=15cm]{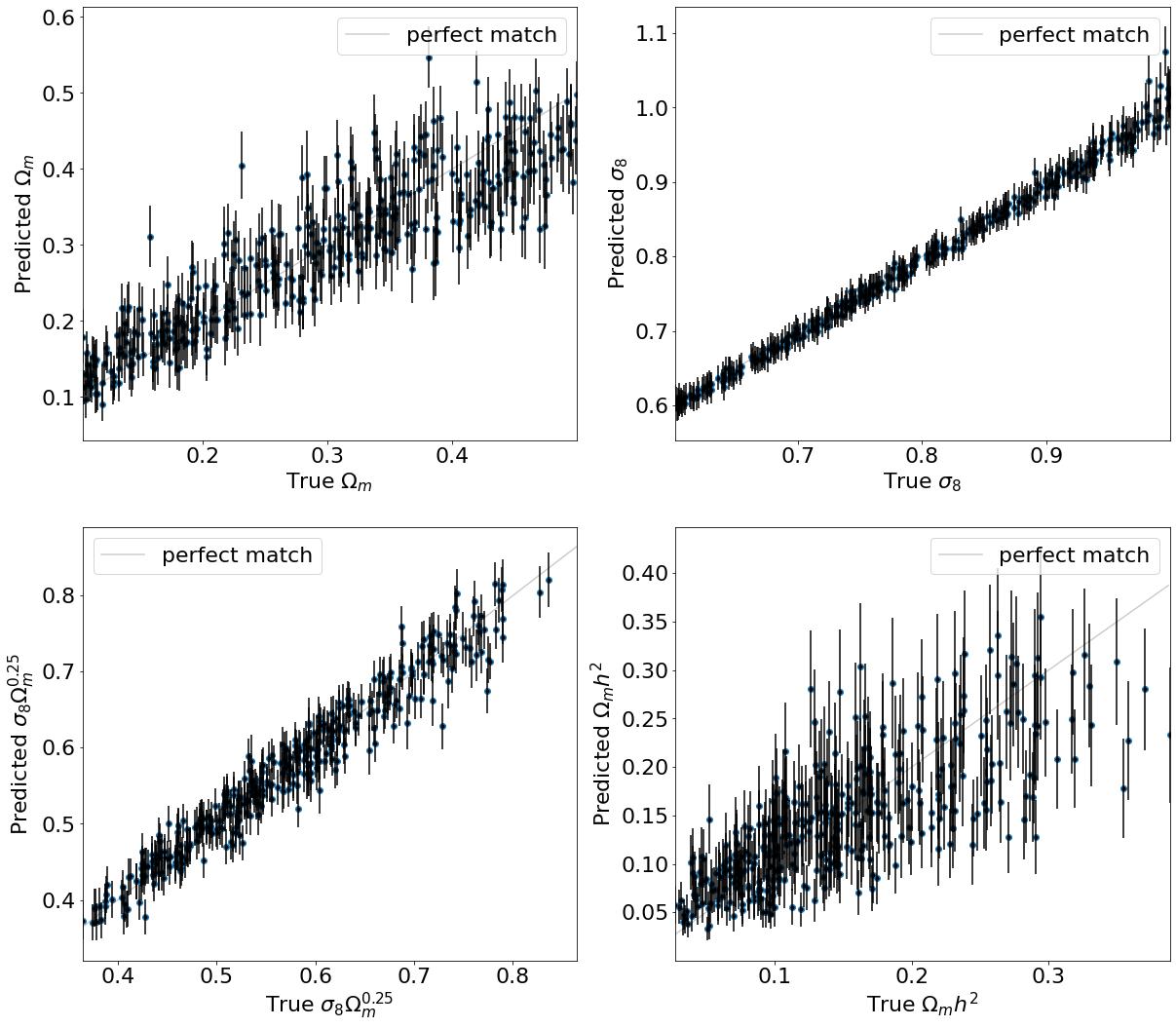}
\end{center}
\caption{ Plots of True vs Predicted values provided by the best experiment  VBNNs, for  $\Omega_m$, $\sigma_8$, and some derivative  parameters. Points are the mean of the predicted distributions, and error bars stand for the heteroscedastic uncertainty associated to  epistemic plus  aleatoric uncertainty at $1\sigma$.  }\label{fig45lines}
\end{figure}
Followed by VBBNs, we have the FlipoutBNNs, however, although this approach yields good cosmological parameter estimation, it understimates their uncertainties.
Therefore, VBNNs avoids indeed the application of an extra post training step in the Machine Learning pipeline related to  calibration. 
Notice that in all experiments, $h$ becomes hardly predicted for all model. Figure~\ref{fig45lines} displays the predicted against true values for $\Omega_m$, $\omega_m$ (instead of $h$), $\sigma_8$ and the degeneracy direction defined as  $\sigma_8\Omega_m^{0.25}$. Error bars report the epistemic plus aleatoric uncertainties predicted by  BNNs, which illustrates the advantages of these probabilistic models where the certainty prediction of the model is captured instead of traditional DNNs where only point estimates are present. This uncertainty was taken from the diagonal part of the covariance matrix. 
\subsection{Calibration metrics}
\begin{figure}[h!]
\begin{center}
\includegraphics[width=15cm]{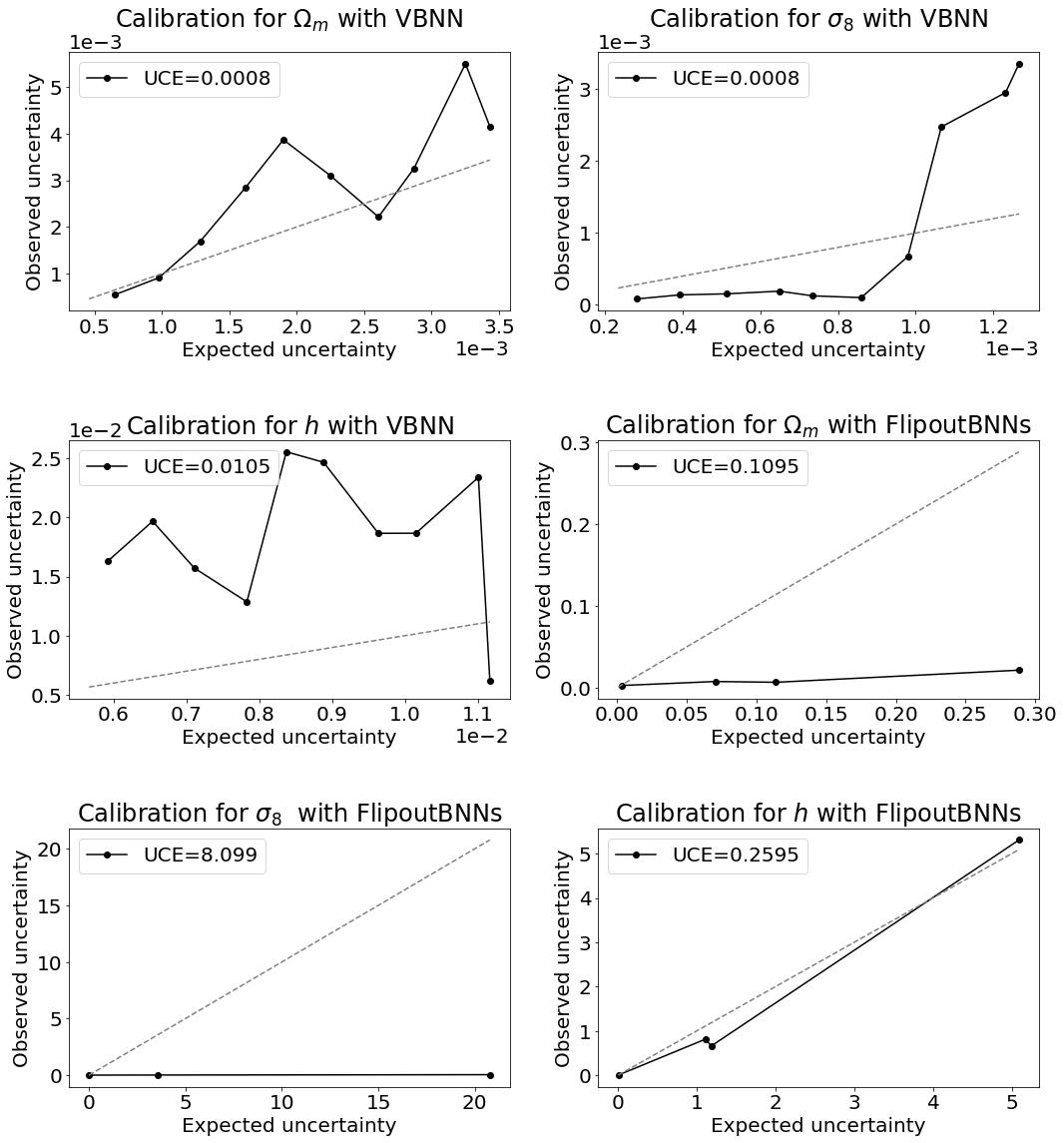}
\end{center}
\caption{ Calibration diagrams for the best experiments, VBNNs and FlipoutBNNs. The lower is the UCE value, the higher  is the calibration of the model. Dashes lines stand for the perfect calibration, so, the discrepancy to this identity curve reveals miscalibration. }\label{figcal}
\end{figure}

In figure~\ref{figcal}, we analyze the quality of our uncertainty measurement using calibration diagrams. We show the predicted uncertainty vs observed uncertainty from our model on the Test dataset.
 Better performing uncertainty estimates should correlate more accurately with the dashed lines. We can see that estimating uncertainty from VBNNs reflect better the real uncertainty. Furthermore,  the scale for VBNNs is two orders of magnitude lower than FlipoutBNN, which also implies how reliable is this models according to their predictions. Notice that the even if we partitioned the variance into $K=10$ bins with equal width,  FlipoutBNNs and sBNNs yield underestimate uncertainties (many examples concentrates in lower bin values), for this reason we see that while  VBNNs supply all ten samples in the calibration plots, for the others we have just 3-4 of them.  Next, we employed the $\sigma$-scaling methodology for calibrating the FlipoutBNNs predictions~\citep{laves2020wellcalibrated}. For doing so, we optimize uniquely the loss function described in  Eq.~\ref{nll} where all parameters related to the BNNs where frozen, i.e., the only trainable parameter was $s$. After training, we got $s\sim 0.723$,  reducing UCE only up to $10\%$, and the number of samples in the calibration diagrams enlarged to 4-5. This minor performance enhancement  means that $\sigma$-scaling is not suitable to calibrate all BNNs, and alternative re-calibration techniques must be taken into account in order to build reliable intervals. At this point, we have noticed the advantages of working with methods that  leading with   networks  already well-calibrated after the training step~\citep{Hort_a_2020a}.    
\subsection{Joint analysis for Cosmological parameters}
\begin{figure}[h!]
\begin{center}
\includegraphics[width=15cm]{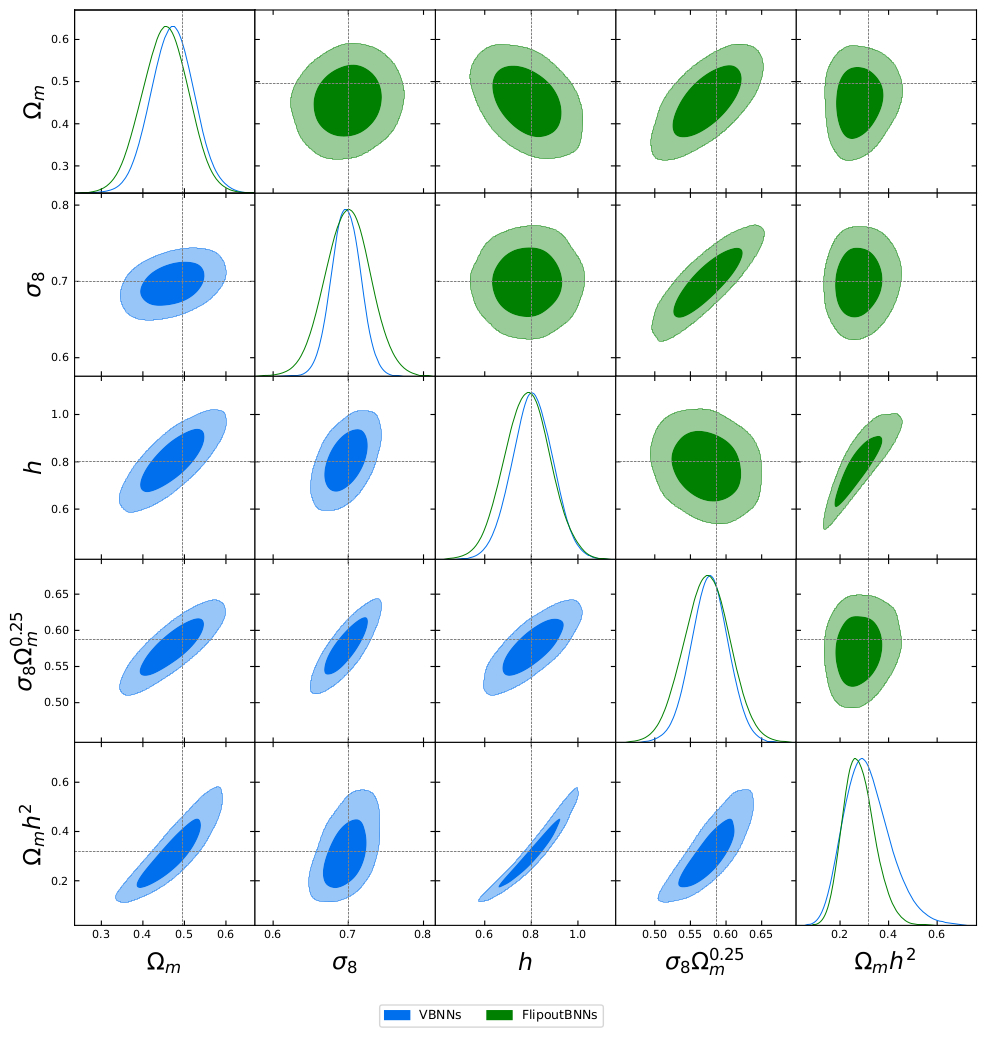}
\end{center}
\caption{68\% and 95\% parameter constraint contours from one example of Quijote test
dataset using VBNNs and FlipoutBNNs. The diagonal plots are the marginalized parameter constraints, the
dashed lines stand for the the true values. This plot was made using Getdist~\citep{Lewis:2019xzd}.}\label{figtp}
\end{figure}

\begin{table}[]
    \centering
  \begin{tabular} { l  c c c}

 Parameter &  95\% limits VBNNs & 95\% limits FlipoutBNNs & True Value \\[0.1cm]
\hline
{\boldmath$\Omega_m       $} & $0.47^{+0.10}_{-0.10}      $ & $0.45^{+0.11}_{-0.11}      $ & 0.495\\[0.1cm]

{\boldmath$\sigma_8       $} & $0.697^{+0.038}_{-0.038}   $ & $0.699^{+0.059}_{-0.060}$   & 0.699\\[0.1cm]

{\boldmath$h              $} & $0.81^{+0.17}_{-0.17}      $  & $0.78^{+0.20}_{-0.19}      $ & 0.800\\[0.1cm]

{\boldmath$\sigma_8\Omega_m^{0.25}   $} & $0.577^{+0.051}_{-0.052}   $  & $0.573^{+0.063}_{-0.064}   $ & 0.587\\[0.1cm]

{\boldmath$\Omega_m h^{2}            $} & $0.31^{+0.19}_{-0.18}      $  & $0.573^{+0.063}_{-0.064}   $ & 0.317\\[0.1cm]
\hline
\end{tabular}
    \caption{Parameter 95\% intervals taken from the parameter constraint contours (figure~\ref{figtp}) from one example of Quijote test dataset using VBNN and FlipoutBNN.}
    \label{tabcp}
\end{table}
In order to show the parameter intervals
and contours from the N-body simulations, we choose randomly an example from the test set
with true values shown in table~\ref{tabcp}.  The two-dimensional posterior distribution of the cosmological parameters are shown in figure~\ref{figtp} and the parameter 95\% intervals are reported in table~\ref{tabcp}. We can observe that VBNNs provides considerably tighter and well constraints on all parameters with respect to the sBNNs~\cite{2112.11865}. Most important, this technique offers also the correlation among parameters and the measurement about how reliable the model in their predictions.
\section{Conclusions }\label{conlcu}
N-body simulations offer one of the most powerful ways to understand the initial conditions of the Universe and improve our knowledge on fundamental physics. In this paper we  used QUIJOTE dataset, in order to show how convolutional DNNs capture non-Gaussian patters without requiring a specifying the summary statistic (such as PS). Additionally, we have show how we can build probabilistic DNNs to obtain uncertainties which account for the reliability in their predictions. One of the main goals of this paper was also reporting how improves these BNNs when we integrate  them with techniques such as a Multiplicative normalizing flows to enhance the  variational posterior complexity. We found that VBNNs not only provides considerably tighter and well constraints on all cosmological parameters as we observed in figure~\ref{figtp}, but also yields with well-calibrated estimate uncertainties as it was shown in figure~\ref{figcal}. Nevertheless, some limitations in this research includes simple prior assumptions  (mean-field approximations),  lower resolution in the simulations, and absence of additional calibration techniques. These restrictions will be analysed in detail in a future paper.

\section*{Acknowledgments}
 This paper is based upon work supported by the Google Cloud Research
Credits program with the award GCP19980904.\\
 Leonardo Castañeda was supported by patrimonio autónomo  fondo Nacional de financiamiento para la ciencia y la tecnología y la innovacion Francisco José de Caldas (Minciencias Colombia) grant No 110685269447 RC-80740-465-2020 projects 69723. H. J. Hort\'ua  acknowledges the support from cr\'editos educaci\'on de doctorados nacionales y en el exterior- colciencias, and  the grant provided by the Google Cloud Research
Credits program.


\bibliographystyle{Frontiers-Vancouver} 
\bibliography{test}





\end{document}